# Strain driven sequential magnetic transitions in strained GdTiO$_3$ on compressive substrates: a first-principles study

Li-Juan Yang,[1,2] Ya-Kui Weng,[2] Hui-Min Zhang,[2] and Shuai Dong[2†]

[1]*Department of Basic Courses, Suqian College, Suqian 223800, China*

[2]*Department of Physics, Southeast University, Nanjing 211189, China*

**Abstract:** The compressive strain effects on the magnetic ground state and electronic structure of strained GdTiO$_3$ have been studied by the first-principles method. Different from the congeneric YTiO$_3$ and LaTiO$_3$ cases both of which becomes the A-type antiferromagnetism on the (001) LaAlO$_3$ substrate despite their contrastive magnetism, the ground state of strained GdTiO$_3$ on the LaAlO$_3$ substrate changes from the original ferromagnetism to G-type antiferromagnetim, instead of the A-type one although Gd$^{3+}$ is between Y$^{3+}$ and La$^{3+}$. Only when the in-plane compressive strain is large enough, e.g. on the (001) YAlO$_3$ substrate, the ground state finally becomes the A-type one. The band structure calculation shows that these compressive strained GdTiO$_3$ remain insulating, although the band gap changes a little in these strained GdTiO$_3$.

---

† All correspondence should be addressed to S.D.: sdong@seu.edu.cn

## I. Introduction

Rare earth titanates ($R$TiO$_3$) are typical Mott insulators with emergent physical properties due to the delicate coupling of electron, lattice, spin, and orbital degrees of freedom. With a single electron occupying Ti's t$_{2g}$ orbitals, $R$TiO$_3$ show magnetic orders, which changes from the ferromagnetism (FM) to G-type antiferromagnetism (G-AFM) with increasing size of $R^{3+}$ [1-6]. The underlying physical mechanism is due to the orthorhombic GdFeO$_3$-type distortion, namely the magnetic ground state is closely coupled with the Ti-O bond angles and bond lengths [1,7-9]. As shown in Fig. 1, with small GdFeO$_3$-type distortions, $R$TiO$_3$ bulks exhibit the G-AFM order ($R$: from La to Sm), while with large distortions, $R$TiO$_3$ bulks show the FM order ($R$: from Gd to Y).

One must note that the lattice distortions of perovskite oxides can be tuned by not only the chemical pressure, but also the epitaxial strain/stress when grown as thin films [10-13]. In addition, thin films are always the necessary way to integrated devices. Therefore, to control the magnetism of $R$TiO$_3$ in thin films is a physical interesting topic and may be technologically important. According to previous studies [14, 15], the A-type antiferromagnetism (A-AFM) was predicted to be the ground state for both LaTiO$_3$ and YTiO$_3$ on the compressive (001) LaAlO$_3$ substrate despite their different original magnetism. Since LaTiO$_3$ and YTiO$_3$ are two end materials of the $R$TiO$_3$ family, it is natural to ask whether every $R$TiO$_3$ becomes A-AFM when grown on the (001) LaAlO$_3$ substrate? Note for those materials locating close to the FM/G-AFM phase boundary, e.g. GdTiO$_3$, the ground states may be even more sensitive to external pressure comparing with the end-cases, and thus significant strain effects may be expected.

In this work, the effects of compressive strain on magnetic and electronic structures of strained GdTiO$_3$ have been studied based on the first-principles theory. Our density functional theory (DFT) calculation predicts that a FM to G-AFM (instead of the A-AFM) phase transition, can be realized by the in-plane compressive strain when using the (001) LaAlO$_3$ substrate, different from the two end $R$TiO$_3$'s. The expected A-AFM can be stablized on the extremely compressive (001) YAlO$_3$ substrates.

## II. Model and method

GdTiO$_3$ bulk has an orthorhombic structure (space group *Pbnm*) with lattice constants of *a*=5.403 Å, *b*=5.7009 Å, and *c*=7.6739 Å [8]. Such a primitive unit cell consists of four formula units. In the following, the in-plane compressive strain is imposed by fixing the lattice constants along both the *a*- and *b*-axis to match the substrates, e.g 5.366 Å on the (001) LaAlO$_3$. Comparing with previous studied YTiO$_3$ and LaTiO$_3$ cases, the average in-plane lattice mismatch with the (001) LaAlO$_3$ substrate is modest (~3.4%), larger than YTiO$_3$ (~3%), but smaller than that of LaTiO$_3$ (~4.9%).

Our DFT calculations are done using the Vienna *ab initio* Simulation

Package (VASP) [16, 17] within the generalized gradient approximation plus $U$ (GGA+$U$) method [18-20]. The 4$f$ electrons of Gd are not included in valence states since the $f$ orbitals are not well treated in VASP. The physical consideration is that Gd's magnetic order only occurs in low temperatures, much below Ti's magnetic ordering temperature. Thus, here we only deal with the magnetism of Ti and leave the low-temperature Gd-Ti coupling to future works. All calculations, including the lattice relaxation and static computations, have been done with the Hubbard $U_{\text{eff}}$=$U$-$J$=3.2 eV by default on the $d$-orbitals of Ti ion using the Dudarev implementation [21]. The optimization and electronic self-consist iterations are performed using the plane-wave cutoff of 550 eV and a 7×7×5 Monkhorst-Pack $k$-point mesh centered at $\Gamma$ point in combination with the tetrahedron method. The inner atomic positions as well as the lattice constants ($a$-$b$-$c$ for the unstrained bulk and $c$ for the strained case) are fully optimized as the Hellman-Feynman forces are converged to be less than 10 meV/Å.

**III. Results and discussion**

First, the ground state of GdTiO$_3$ bulk is checked. Using the fully optimized crystal structure, four magnetic orders: FM, A-AFM, C-AFM, and G-AFM have been calculated to compare the energies. As shown in Table I, the FM order has the lowest energy, implying the ground state, in consistent with the experimental result [1,8]. According to Table I, other magnetic orders' energies (per Ti) are higher than the FM one: 2.8 meV higher for A-type AFM, 8.0 meV higher for C-type AFM, and 9.9 meV higher for G-type AFM. The fully relaxed lattice gives: $a$=5.434 Å, $b$=5.809 Å, $c$=7.765 Å, very close to the experimental data. The FM ground state has also been confirmed within a wide range of $U_{eff}$ from 0 eV to 4 eV. When $U_{eff}$=3.2 eV, the calculated band gap is 1.68 eV from the band structure, The calculated magnetic moment is 0.88 $\mu_B$/Ti. If the Hubbard $U_{eff}$ is not included, the pure GGA calculation will get a metallic DOS, implying the Coulomb repulsion between 3$d$ electrons drives GdTiO$_3$ to be a Mott insulator.

Subsequently, our calculations are done with the epitaxial strain. By fixing the in-plane lattice constants to fit the (001) LaAlO$_3$ substrate as stated before, the equilibrium lattice constant along the $c$-axis is searched from 7.0 Å to 9.0 Å. For each $c$-value, the internal atomic positions are relaxed with magnetism to obtain the optimal crystal structure with the lowest energies. According to Fig. 2(a), in the whole calculated region, the G-AFM state is the lowest one in energy than all other states. This result is quite different from the previous studied YTiO$_3$ and LaTiO$_3$ cases [14, 15], both of which show an A-AFM ground state despite their original magnetic orders. The C-AFM state is the first excited state, which is also nontrivial different from the strained YTiO$_3$ and LaTiO$_3$. The relaxed lattice constants along the $c$-axis are both 8.28 Å for the C-AFM and G-AFM. As summarized in Table II, the C-AFM state is 8.4 meV/Ti higher in energy than the G-AFM at $c$=8.28 Å, while others are even higher. As shown in Fig. 2(b), the energy difference between the G-AFM and C-AFM states does not change too much by varying the *c*-axis lattice constant from 8.2 Å to 8.3 Å, suggesting the result is not sensitive to the *c*-axis lattice constant near the optimal value. Note that the energy difference between the ground state and the first excited state is only 1.9 meV/Ti in bulk GdTiO$_3$. Thus, the G-AFM state should be very robust under such a compressive strain. To further check the reliability, the energy difference between the G-AFM and C-AFM states is calculated with different $U_{\text{eff}}$ from 0 eV to 5 eV stepped by 1 eV, which varies from -17.1 meV to -4.8 meV, as shown in Fig. 2(c). All these data illustrate that the G-AFM state is very promising for strained GdTiO$_3$ on the (001) LaAlO$_3$ substrate.

As mentioned before, the magnetic ground state of *R*TiO$_3$ depends on the GdFeO$_3$-type distortion. Thus, the Ti-O-Ti bond angles and bond length in strained GdTiO$_3$ on the LaAlO$_3$ substrate (G-AFM) are compared in Table III, together with the bulk's values (FM), which may help us to understand the underlying physical mechanism. The bond angle along the *c*-axis increases by 2.6º but the one in the *ab*-plane decreases by 4.7º. The increased bond angle and bond length along the *c*-axis certainly prefer the antiferromagnetic coupling, according to the bulk phase diagram and previous experience of YTiO$_3$ and LaTiO$_3$. However, the antiferromagnetic coupling, accompanied by the decreased in-plane bond angle and bond length, is exotic, which can not be simply understood from the bond angle/length effect and suggesting a more complex physical mechanism involved.

It is important to check the density of states (DOS) of GdTiO$_3$ under strain to study the conductance accompanying the magnetic transition, as shown in Fig. 3(a). The GGA+$U$ calculation found the gap as 1.76 eV for strained GdTiO$_3$ on (001) LaAlO$_3$ substrate, which is a little larger than the bulk value 1.68 eV. In Fig. 3(b), this band gap is plotted as a function of $U_{\text{eff}}$ from 0 eV (pure GGA) to 5 eV. As in the bulk case, the pure GGA calculation gives a metallic behavior, while the band gap increases almost linearly with $U_{\text{eff}}$, suggesting a Mott insulator fact for strained GdTiO$_3$ on LaAlO$_3$ substrate.

The physical reason for such strain-driven G-AFM can be understood as following. According to the bulk phase diagram, GdTiO$_3$ locates very close to the FM/G-AFM phase boundary, implying proximate energy between these contrastive phase and strong quantum fluctuation. According to the bulk's phase diagram and previous studies, the compressive strain suppresses the FM order and favors the A-AFM. However, when the strain is modest, the energy of A-AFM remains not low enough. Alternatively, the G-AFM beats the FM due to its antiferromagnetic correlation. In other word, the intrinsic quantum competition in GdTiO$_3$ is the dominant driven force, while the compressive strain plays as an assistant role. In contrast, for previous studied LaTiO$_3$ and YTiO$_3$ cases, both of them locate far from the FM/G-AFM boundary, implying a weak quantum fluctuation. Therefore, the strain effect is almost the pure driven force for phase transition, giving rise to the expected A-AFM phase under compressive strain. Of course, the "unexpected" G-AFM in strained GdTiO$_3$ predicted in the present calculation desires following experiments to verify it and further theortical studies to understand the underlying mechanism.

In this sense, it is interesting to expect that a stronger compressive strain may overcome the intrinsic competition and play as the main role. To confirm this speculation, we fix the in-plane lattice constants of GdTiO$_3$ to 5.1377 Å and 5.2736 Å, fitting even smaller (001) YAlO$_3$ substrate. For this substrate, the biaxial strain is 6.2% for the YAlO$_3$ substrate, larger than the LaAlO$_3$ one (~3.4%). Also the internal atomic positions and the lattice constant along the $c$-axis are re-relaxed with various magnetic orders to search the optimized structure and ground state. As shown in Table IV, now the A-AFM state is the ground state with the lowest energy, and the FM state is the first excited state, in consistent with previous studied compressive YTiO$_3$ and LaTiO$_3$. The relaxed lattice constant along the $c$-axis is 8.63 Å for the A-AFM state, longer than above case on the LaAlO$_3$ substrate. The energy difference between the A-AFM state and FM state is also calculated by varying $U_{eff}$ to further check the reliability. As shown in Fig. 4(b), the energy difference between them is always negative. All these results suggest that the A-AFM state is a robust state for the strained GdTiO$_3$ on extremely compressive YTiO$_3$ substrate. The DOS is also checked, as shown in Fig. 4(a). The gap is 1.64 eV, remaining a Mott insulator.

At last, the band structures for GdTiO$_3$ bulk and strained GdTiO$_3$ (G-AFM and A-AFM) are also calculated, as shown in Fig. 5. First, these band structures again prove the ground states for GdTiO$_3$ bulk and strained GdTiO$_3$ are all Mott insulators. Second, the topmost valence bands close to the Fermi level are from the $t_{2g}$ orbitals of Ti, which are nearly degenerated. Third, the lowest conducting bands are also from the $t_{2g}$ orbitals of Ti, which are more than 1 eV higher than the occupied bands. Such a large splitting of $t_{2g}$ levels is due to the orbital ordering associated with the Jahn-Teller distortion, as well as the inter-orbital Hubbard repulsion. Fourth, the bandwidth of these occupied $t_{2g}$ levels is very narrow, or namely nearly flat, implying that these occupied states are almost isolated, with weak kinetic energy. Even though, the bandwidth changes when GdTiO$_3$ is strained: shrinking from 0.445 eV (FM, bulk) to 0.248 eV (G-AFM, on LaAlO$_3$ substrate), then to 0.412 eV (A-AFM, on YAlO$_3$ substrate). This change, together with the change of bands above the Fermi level, is the driving force for phase competition in such a multi-orbital correlated system.

In summary, the magnetic orders of strained GdTiO$_3$ on two compressive substrates have been studied based on density functional theory calculations. A phase transition from the original ferromagnetism in bulk to G-type antiferromagnetism has been found on the (001) LaAlO$_3$ substrate, which is different from the previous studied LaTiO$_3$ and YTiO$_3$. The underlying physics for such a unique transition is the intrinsic phase competition which dominates the modest strain. Only when the compressive strain is very large, e.g. using the (001) YAlO$_3$ substrate, the ground state changes to the A-type antiferromagnetism as compressive strain on LaTiO$_3$ and YTiO$_3$. Furthermore, these strain-induced magnetic transition will not change the insulating behavior of GdTiO$_3$ but only tune the band gap a little.

## Acknowledgments

Work was supported by the Natural Science Foundation of China (grant Nos. 51322206 and 11274060) and the 973 Projects of China (grant No. 2011CB922101).

# Figures

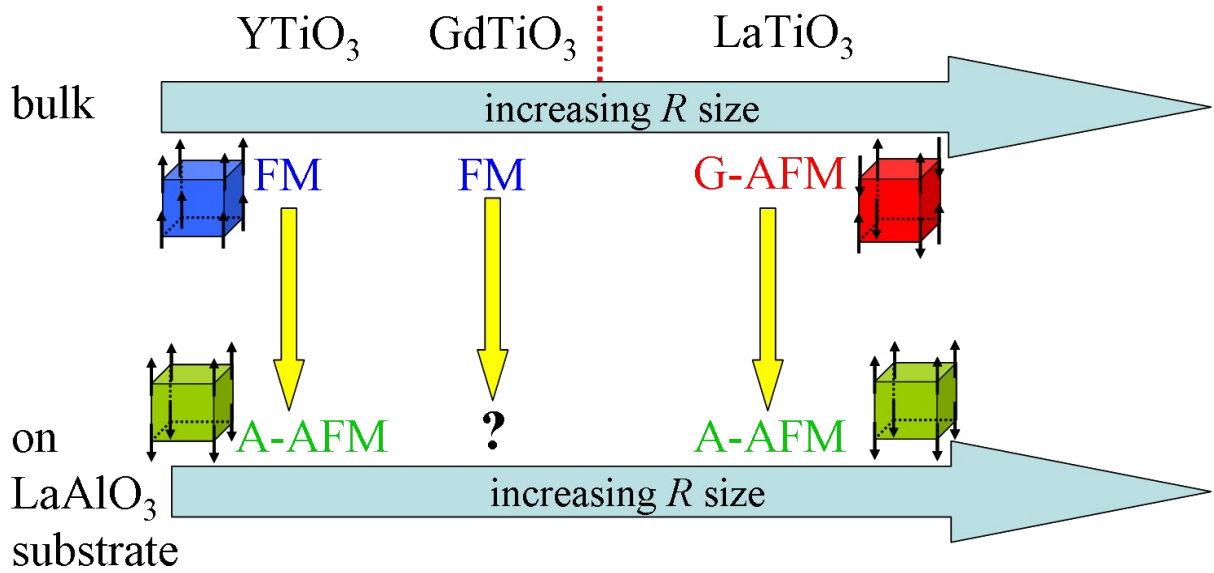

FIG. 1. (Colour online) Sketch of compressive strain effect in $R$TiO$_3$. GdTiO$_3$ locates in the middle of two end materials YTiO$_3$ and LaTiO$_3$, close to the phase boundary of FM and G-AFM. Our previous DFT calculations predicted the strain induced A-AFM for both YTiO$_3$ and LaTiO$_3$ on the (001) LaAlO$_3$ substrate.

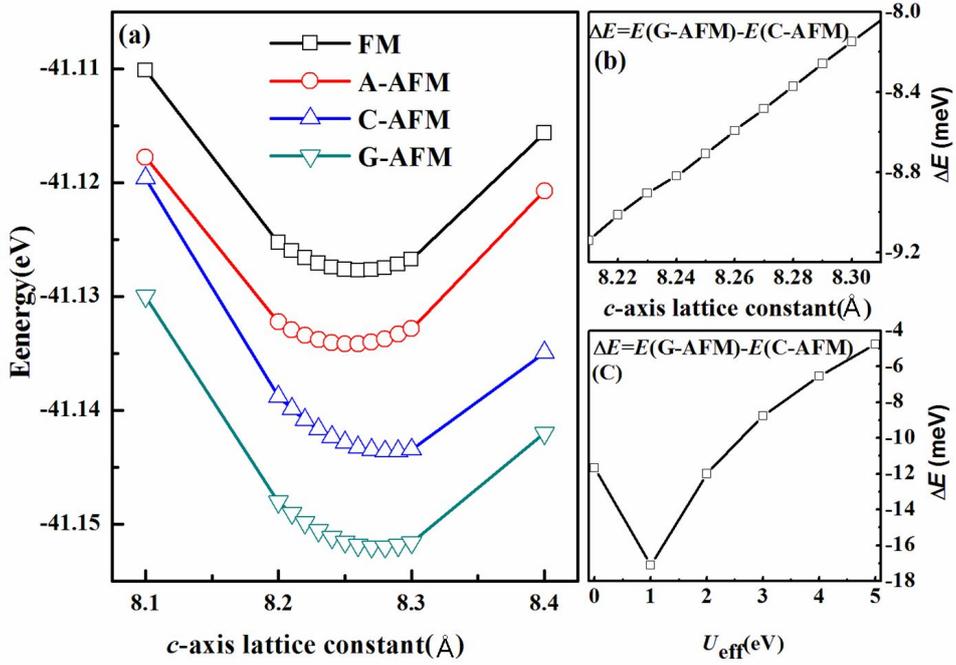

FIG. 2. (Colour online) (a) Energies (per Ti) for various magnetic orders as a function of the *c*-axis lattice constant. (b) The energy difference (per Ti) between the G-AFM and C-AFM as a function of the *c*-axis lattice constant near the optimal *c*-axis value. (c) The energy difference (per Ti) between the G-AFM and C-AFM as a function of the Hubbard $U_{\text{eff}}$. The $U=0$ case is calculated using the pure GGA method, which shows a little abnormal comparing with the GGA+$U$ method .

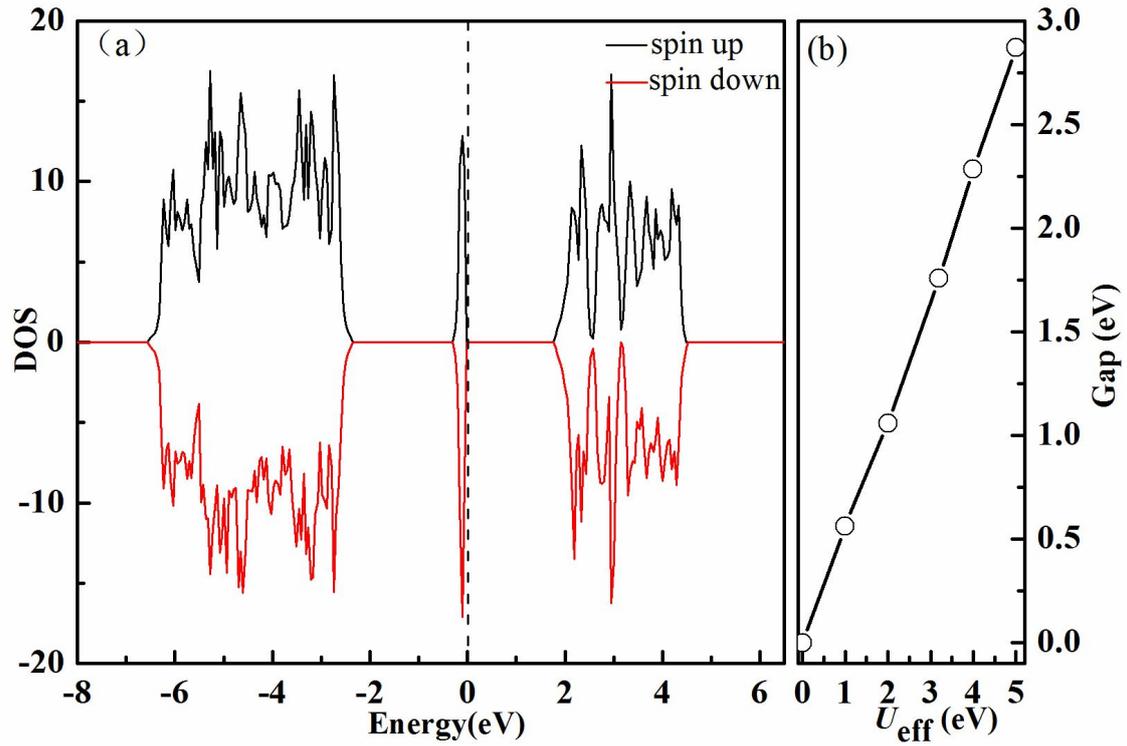

FIG. 3. (Colour online) (a) Total DOS of strained GdTiO$_3$ on LaAlO$_3$ substrate. The Fermi energy is positioned at zero. (b) The band gap as a function of $U_{\text{eff}}$. The critical $U_{\text{eff}}$ value to induce a gap is estimated to be 0.4 eV by extrapolation.

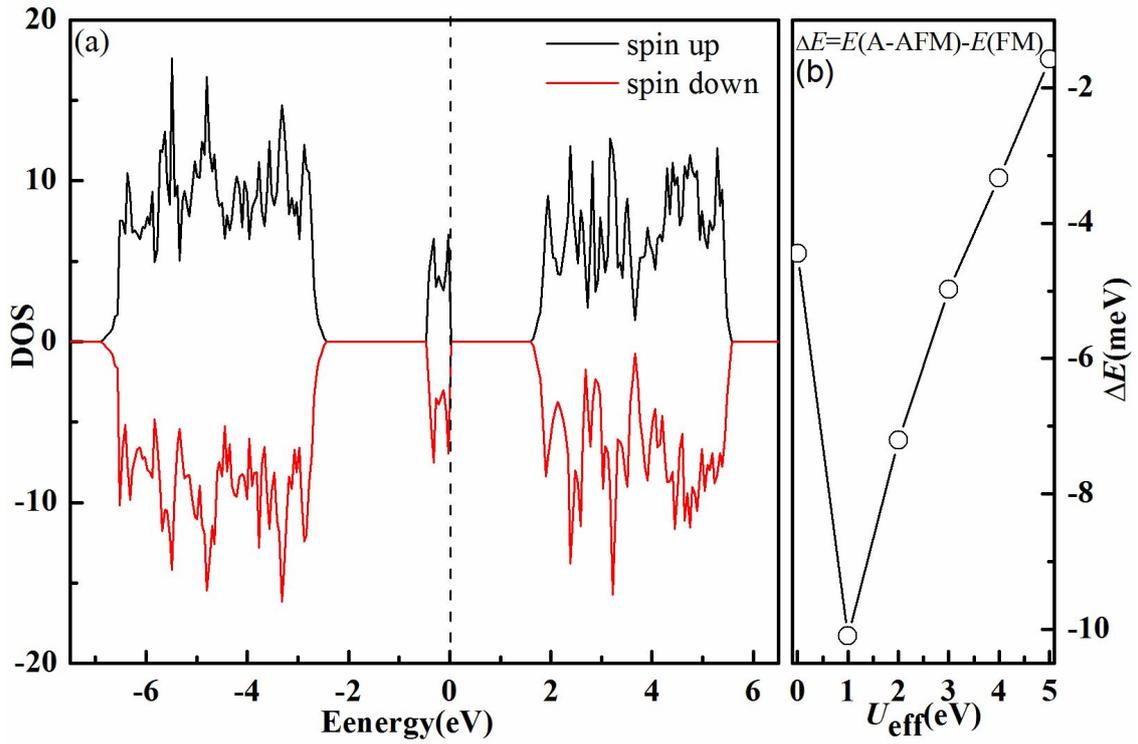

FIG. 4. (Colour online) (a) Total DOS of strained GdTiO$_3$ on YAlO$_3$ substrate. The Fermi energy is positioned at zero. (b) The energy difference (per Ti) between the A-AFM and FM as a function of the Hubbard $U_{\text{eff}}$.

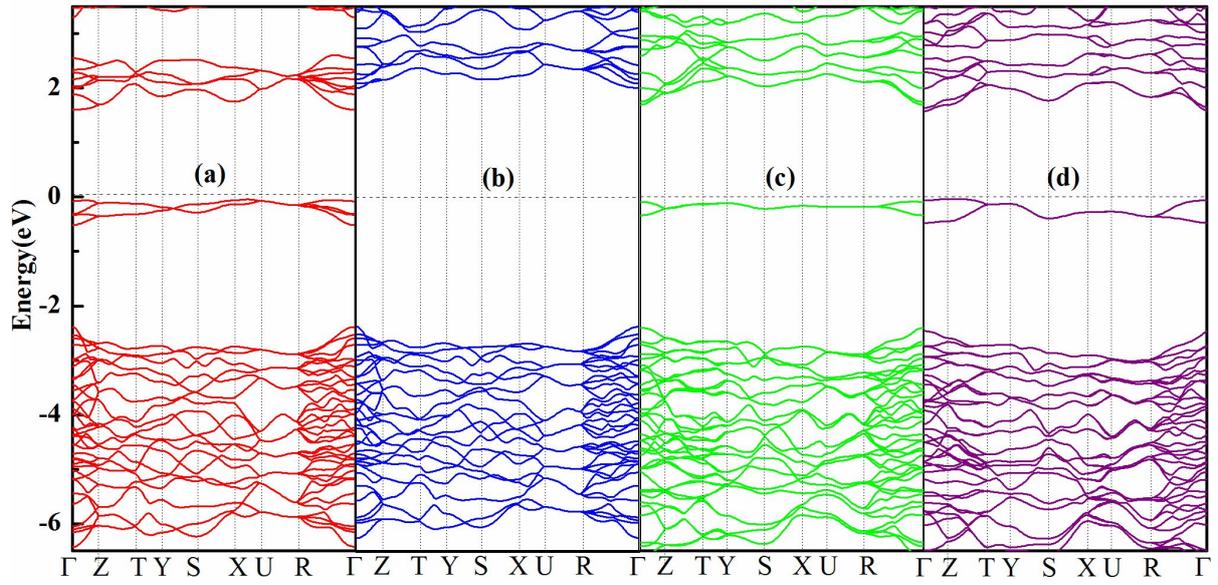

FIG. 5. (Colour online) The band structure of GdTiO$_3$. (a-b) FM bulk. (a) Spin up. (b) Spin down. (c) G-AFM state on LaAlO$_3$ substrate. (d) A-AFM state on YAlO$_3$ substrate.

# Tables

TABLE I. The energy difference $\Delta E$/meV (per Ti) between various magnetic states for GdTiO$_3$ bulk: $E(X)$-$E$(FM), and the corresponding local magnetic moments $M$ are also shown in unit of $\mu_B$/Ti.

| Magnetic order | FM | A-AFM | C-AFM | G-AFM |
|---|---|---|---|---|
| $\Delta E$ | 0 | +2.8 | +8.0 | +9.9 |
| $M$ | 0.878 | 0.869 | 0.848 | 0.842 |

TABLE II. The energy difference $\Delta E$/ meV (per Ti) between various magnetic states for strained GdTiO$_3$ on LaAlO$_3$ substrate: $E(X)$-$E$(G-AFM), the corresponding value of $c$/Å and local magnetic moments $M$/$\mu_B$ (Per Ti) are also shown.

| Magnetic order | FM | A-AFM | C-AFM | G-AFM |
|---|---|---|---|---|
| $\Delta E$ | +23.6 | +17.4 | +8.4 | 0 |
| $c$ | 8.26 | 8.25 | 8.28 | 8.28 |
| $M$ | 0.875 | 0.870 | 0.859 | 0.856 |

TABLE III. Calculated bond angles and bond lengths in the ab-plane and along c-axis of strained GdTiO$_3$ on the LaAlO$_3$ substrate, compared with bulk GdTiO$_3$.

| Ti-O-Ti bond angle<br>Ti-O bond length | Strained GdTiO$_3$<br>(G-AFM) | Bulk GdTiO$_3$<br>(FM) |
|---|---|---|
| ab-plane | 140.7° | 145.4° |
|  | 2.052 Å/1.977 Å | 2.082 Å/2.032 Å |
| c-axis | 145.6° | 143.3° |
|  | 2.167 Å | 2.021 Å |

TABLE IV. The energy difference $\Delta E$/ meV (per Ti) between various magnetic states for strained GdTiO$_3$ on YAlO$_3$ substrate: $E(X)$-$E$(A-AFM), the corresponding value of $c$/Å and local magnetic moments $M/\mu_B$ (Per Ti) are also shown.

| Magnetic order | FM | A-AFM | C-AFM | G-AFM |
|---|---|---|---|---|
| $\Delta E$ | +4.6 | 0 | +9.5 | +5.2 |
| $c$ | 8.64 | 8.63 | 8.64 | 8.63 |
| $M$ | 0.880 | 0.874 | 0.856 | 0.854 |